\documentstyle[preprint,prd,aps]{revtex}
\begin{document}
\draft
\preprint{
\vbox{
\halign{&##\hfil\cr
	& NUHEP-TH-95-9 \cr
	& hep-ph/9508373 \cr
	& August 1995 \cr}}
}
\title{Signature for Color-Octet Production of $J/\psi$ \\
	in $e^+ e^-$ Annihilation}

\author{Eric Braaten}
\address{Department of Physics, Ohio State University, Columbus OH 43210}

\author{Yu-Qi Chen}
\address{Department of Physics and Astronomy, Northwestern University,
	Evanston, IL 60208}

\maketitle

\begin{abstract}
The inclusive cross section for the production of $J/\psi$ in $e^+ e^-$
annihilation is studied using a recently developed factorization formalism.
In addition to the conventional color-singlet contribution, we include
contributions from $c \bar c$ pairs that are produced at short distances
in color-octet states.  The color-octet contributions are suppressed by
the smaller probability for the color-octet $c \bar c$ pair to evolve
into a state that includes a $\psi$, but they may dominate near the upper
endpoint of the $\psi$ energy spectrum. The signature for the color-octet
contribution is a dramatic change in the angular distribution of the $\psi$
near the endpoint.
\end{abstract}

\vfill \eject

\narrowtext

Charmonium is a bound state of a charm quark and antiquark held together
by the strong force of quantum chromodynamics (QCD). Because QCD confines
color charges, the quark and antiquark must be in a singlet state with
respect to color. The production of charmonium requires the creation of
a $c \bar c$ pair with energy greater than $2 m_c$, where $m_c$ is the
mass of the charm quark. Since the QCD coupling constant is relatively
weak at the scale $2 m_c$, the production of charmonium should be amenable
to theoretical analysis using perturbation theory in $\alpha_s$.
Most theoretical predictions of charmonium production have been based on
the ``color-singlet model'' \cite{schuler}.  In this model, the $c \bar c$
pair that forms charmonium is assumed to be produced in a color-singlet
state by parton collisions whose cross sections can be calculated using
perturbation theory.  The probability that a $c \bar c$ pair
produced in a color-octet state will bind to form charmonium is assumed to
be negligible.

Recent developments
in both theory and experiment have challenged this simple assumption.
Bodwin, Braaten and Lepage \cite{BBL} have developed a factorization
formalism that allows the systematic calculation of inclusive
cross sections for charmonium states to any order in the
QCD coupling constant $\alpha_s$ and to any order in $v^2$, where $v$
is the typical relative velocity of the charm quark.
This formalism implies that ``color-octet processes'',
in which the $c \bar c$ pair that forms charmonium is produced
at short distances in a color-octet state, must contribute to the
cross section at some level.  The assumption that charmonium production
is always dominated by color-singlet processes has also been
challenged by recent experimental data.
The CDF collaboration has measured the cross sections for prompt
charmonium production at large transverse momentum at the Tevatron \cite{CDF},
using a silicon vertex detector to separate the prompt signal from
the background of charmonium produced by the decay of $B$ hadrons.
At the large values of $p_T$ available at the Tevatron, the dominant
production mechanism is fragmentation \cite{braaten-yuan},
the formation of charmonium within the
jet of a high energy parton.  For the $J^{PC} = 1^{--}$ charmonium states
$J/\psi$ and $\psi'$, the prompt cross sections were measured to be more
than an order of magnitude larger than the predictions of the
color-singlet model, even after including the effects of fragmentation
\cite{bdfm}.

One of the proposed explanations for the CDF data on prompt $\psi$
production is that the large cross section comes from a color-octet
term in the fragmentation function for a gluon to form $\psi$
\cite{braaten-fleming}.  This term corresponds to the gluon forming
a $c \bar c$ pair in a color-octet $^3S_1$ state at short distances
of order $1/m_c$ or smaller,
with the binding of the $c \bar c$ pair into a color-singlet $\psi$
occurring at longer distances.
This explanation reproduces the $p_T$-dependence of the CDF data.
The normalization of the data can be fit by adjusting the value
of a matrix element that measures the probability for the formation
of $\psi$ from a pointlike $c \bar c$ pair in a color-octet
$^3S_1$ state \cite{cho}.
This matrix element is found to be much smaller than the corresponding
matrix element in the color-singlet model, which
measures the probability for the formation
of $\psi$ from a pointlike $c \bar c$ pair in a color-singlet $^3S_1$ state.
In spite of the fact that the color-octet matrix element is intrinsically
smaller, the color-octet term can dominate over the color-singlet
term in the gluon fragmentation function because the color-singlet term
is suppressed by a short-distance factor of $\alpha_s^2(m_c)$.

In order to establish that the color-octet mechanism plays a
significant role in the production of charmonium, it is necessary
to find other production processes in which it dominates over the
conventional color-singlet mechanisms.
In this Letter, we point out that there is a dramatic signal
for color-octet processes in inclusive
$\psi$ production in $e^+ e^-$ annihilation.
The color-octet terms are suppressed relative to the color-singlet
term over most regions of phase space, but they are enhanced
near the upper endpoint of the $\psi$ energy distribution.
The distribution of the angle $\theta$
between the 3-momentum of the $\psi$ and the beam axis has the form
\begin{equation}
{d \sigma \over dE d \cos \theta}
\; =  \; S(E) \left( 1 \;+\; A(E) \cos^2 \theta \right) \;.
\end{equation}
where $E$ is the energy of the $\psi$ in the center of momentum frame.
The shape of the angular distribution is characterized by
the parameter $A(E)$, which varies with $E$.
The color-singlet model predicts that $A(E)$ should approach
a value close to $-1$ near the upper endpoint of $E$.
We show in this paper that,
if color-octet contributions dominate near the endpoint,
$A(E)$ will change sign and approach a value close to $+1$ at the endpoint.
This change in the sign of $A$ would provide a dramatic signature
for color-octet production mechanisms.

Using the formalism of Ref. \cite{BBL}, the inclusive cross section
for the production of a charmonium state $H$ with momentum $P$ can be
written in a factored form:
\begin{equation}
d \sigma (e^+ e^- \to H(P) + X)
\;=\; \sum_n d {\widehat \sigma} (e^+ e^- \to c \bar c(P,n) + X) \;
	\langle {\cal O}^H_n \rangle,
\label{fact}
\end{equation}
where $d {\widehat \sigma}$ is the inclusive
cross section for producing a $c \bar c$ pair with total
momentum $P$, vanishing relative momentum,
and in a color and angular-momentum state labelled by $n$.
This parton cross section is a short-distance quantity that can be calculated
as a perturbation series in $\alpha_s(m_c)$.
The long-distance factor $\langle {\cal O}^H_n \rangle$
is the matrix element of a local 4-fermion operator in nonrelativistic QCD
which represents the inclusive probability for the formation of the meson
$H$ from the pointlike $c \bar c$ state $n$.
All dependence on the particular quarkonium state $H$
resides in the matrix elements.
The relative magnitudes  of the matrix elements
$\langle {\cal O}^H_n \rangle$ can be estimated using scaling rules
that predict how they scale with the quark mass $m_c$ and with the relative
velocity $v$ of the quark.

In the case of P-wave states, there is a color-octet matrix element
that contributes at the same order in $v^2$ as the leading color-singlet
matrix element.  The color-octet term is needed for perturbative
consistency, because it cancels an infrared
divergence in the color-singlet term \cite{bbly}.
In the case of S-waves, the leading color-octet matrix element
is suppressed by $v^4$ relative to the leading color-singlet matrix element.
It is therefore perturbatively consistent to neglect the color-octet term.
However, the suppression factor $v^4$ is only about 1/10 for charmonium,
so the color-octet term is not necessarily negligible numerically.

The predictions of the color-singlet model for the
inclusive production of $\psi$ in $e^+ e^-$ annihilation
have been explored thoroughly \cite{keung,kuhn,mirkes}.
In this model, the sum (\ref{fact}) is truncated to
a single term whose matrix element is proportional to $|R(0)|^2$,
where $R(0)$ is the radial wavefunction at the origin.
In the notation of Ref.~\cite{BBL}, the matrix element is
$\langle {\cal O}^{\psi}_1(^3S_1) \rangle \approx (9/2 \pi) |R(0)|^2$.
At leading order in $\alpha_s$, the parton process is
$e^+ e^- \to c \bar c g g$, where the $c \bar c$ pair is produced in a
color-singlet $^3S_1$ state with vanishing relative momentum.
The corresponding differential cross section
$d \widehat{\sigma}_1^{(^3S_1)}$ can be obtained from Ref.~\cite{mirkes}
by integrating over the energies of the recoiling gluons.
The complete expression
for the differential cross section is too lengthy to reproduce here,
but it becomes relatively simple at the endpoints of the energy spectrum.
Near the lower endpoint $E_{\rm min} = 2 m_c$, the distribution is
isotropic, corresponding to $A =0$.
The upper endpoint is $E_{\rm max} = (s + 4 m_c^2)/(2 \sqrt{s})$,
where $s$ is the square of the center of mass energy.
At this endpoint, the distribution approaches
\begin{equation}
{d \widehat{\sigma}_1^{(^3S_1)} \over dE d \cos \theta}
\; \longrightarrow \; {128 \pi \alpha^2 \alpha_s^2 \over 243 m_c s^2}
\left( {1+r \over 1-r} \;-\; \cos^2 \theta \right) \;,
\label{csm-max}
\end{equation}
where $r = 4 m_c^2/s$.
For comparison with data, we can identify
$2 m_c$ with the mass $M_\psi$ of the $\psi$.  Thus the color-singlet
model predicts that $A(E)$ should vary from 0 at $E = M_\psi$
to $-(s-M_\psi^2)/(s+M_\psi^2)$
at $E_{\rm max} = (s + M_\psi^2)/(2 \sqrt{s})$.
The prediction for $A(E)$ as a function of the $\psi$ momentum
$p_\psi = \sqrt{E^2 - M_\psi^2}$
at $\sqrt{s} = 10.6 \; {\rm GeV}$ is shown as a solid line in Fig.~1.
The predicted value at the endpoint is $A = -0.84$.

Color-octet contributions to the differential cross section
(\ref{fact}) are suppressed by a factor of $v^4$ from the matrix
element $\langle {\cal O}_n^\psi \rangle$.  They can be competitive
with the color-singlet terms only if there is some enhancement
from the short-distance cross section $d \widehat{\sigma}$.
Such an enhancement does occur in the endpoint region.
At leading order in $\alpha_s$, the color-octet
contribution comes from the parton process $e^+ e^- \to c \bar c g$,
which produces $c \bar c$ pairs only in the endpoint region
and with angular momentum $^1S_0$, $^3P_J$, $^1D_0$, etc.
{}From the velocity-counting rules of Ref.~\cite{BBL},
the dominant matrix elements at this order in $\alpha_s$ are
$\langle {\cal O}^{\psi}_8(^1S_0) \rangle$ and
$\langle {\cal O}^{\psi}_8(^3P_J) \rangle$, $J = 0,1,2$, which are
suppressed by $v^4$ relative to the color-singlet matrix element
$\langle {\cal O}^{\psi}_1(^3S_1) \rangle$.
The matrix element $\langle {\cal O}^{\psi}_8(^1S_0) \rangle$
is proportional to the probability for the formation of a $\psi$
from a pointlike $c \bar c$ pair that is produced in a color-octet
$^1S_0$ state, and similarly for $\langle {\cal O}^{\psi}_8(^3P_J) \rangle$.
The suppression by $v^4$ arises because the formation of the color-singlet
bound state can proceed through a magnetic-dipole
transition in the $^1S_0$ case and through a double electric-dipole
transition in the $^3P_J$ case.
The three P-wave matrix elements are not independent at leading
order in $v^2$, but are related by heavy-quark spin symmetry:
\begin{equation}
\langle {\cal O}^{\psi}_8(^3P_J) \rangle
\;\approx\; (2J+1) \langle {\cal O}^{\psi}_8(^3P_0) \rangle \;.
\end{equation}
The short-distance cross sections corresponding to these matrix elements
can be calculated from the Feynman diagrams for $e^+ e^- \to c \bar c g$:
\begin{eqnarray}
{d \widehat{\sigma}_8^{(^1S_0)} \over dE d \cos \theta}
&=& \delta(E - E_{\rm max}) {16 \pi^2 \alpha^2 \alpha_s \over 9 m_c s^2}
(1-r) (1 + \cos^2\theta) \;,
\label{com-S}
\\
{d \widehat{\sigma}_8^{(^3P_0)} \over dE d \cos \theta}
&=& \delta(E - E_{\rm max}) {32 \pi^2 \alpha^2 \alpha_s \over 27 m_c^3 s^2}
{(1-3r)^2 (1 + \cos^2 \theta) \over 2 (1-r)} \;.
\label{com-P0}
\\{d \widehat{\sigma}_8^{(^3P_1)} \over dE d \cos \theta}
&=& \delta(E - E_{\rm max}) {32 \pi^2 \alpha^2 \alpha_s \over 27 m_c^3 s^2}
{(1+2r) + (1-2r) \cos^2 \theta \over (1-r)} \;.
\label{com-P1}
\\{d \widehat{\sigma}_8^{(^3P_2)} \over dE d \cos \theta}
&=& \delta(E - E_{\rm max}) {32 \pi^2 \alpha^2 \alpha_s \over 27 m_c^3 s^2}
{(1+6r+6r^2) + (1-6r+6r^2) \cos^2 \theta \over 5 (1-r)} \;.
\label{com-P2}
\end{eqnarray}
The delta functions should be interpreted as narrow distributions
in the energy.  In the quarkonium rest frame, the width of the delta
distribution is of order $m_c v^2$.
This is the order of magnitude of the splittings
between the radial and orbital-angular-momentum energy levels of charmonium,
which are roughly 500 MeV.  The effect of boosting to
the $e^+ e^-$ center of momentum frame is to decrease the width
of the delta function by the factor $2 m_c/\sqrt{s}$,
which at CLEO energies is approximately 0.3.  Thus the
delta functions in (\ref{com-S})--(\ref{com-P2}) should be interpreted
as distributions in $E$ whose widths are on the order of 150 MeV.

We now compare the magnitudes of the color-singlet and color-octet
contributions to the cross section.  Away from the endpoint region,
the color-singlet and color-octet parton cross sections
$d \widehat{\sigma}$ are both
of order $\alpha_s^2$.  The color-octet terms are therefore suppressed
by a factor of $v^4$ from the matrix elements.  In the endpoint region,
the parton cross sections (\ref{com-S})--(\ref{com-P2}) for the
color-octet terms are of order $\alpha_s$ and they are enhanced by
a factor of $1/v^2$ from the energy delta functions.
The relative magnitudes of the color-octet terms compared to the
leading color-singlet term  is therefore $v^2/\alpha_s$
in the endpoint region.  Thus the color-octet
terms can be competitive in this region.
They will dominate over the color-singlet term
if the $1/\alpha_s$ enhancement overcomes the $v^2$ suppression.

If we assume that the color-octet terms dominate in the endpoint region,
we obtain a simple and dramatic prediction for the $\psi$ angular distribution.
In the endpoint region, the distribution
should approach a form that corresponds to some linear combination of the
differential cross sections $d \widehat{\sigma}_8^{(^1S_0)}$ and
$\sum_J (2J+1) d \widehat{\sigma}_8^{(^3P_J)}$.
If the $^1S_0$ term dominates,
the angular distribution parameter $A(E)$ should approach
$+1$.  If the $^3P_J$ terms dominate,
then $A(E)$ should approach the value $(3-10r+7r^2)/(3+6r+7r^2)$.
At CLEO energies, this value is approximately $+0.62$.
Thus, regardless of the relative sizes of the matrix elements
$\langle {\cal O}^{\psi}_8(^1S_0) \rangle$ and
$\langle {\cal O}^{\psi}_8(^3P_0) \rangle$, the value of $A$
is predicted to be positive and close to +1.
The predictions from color-octet dominance
of the endpoint region fall in the hatched region in Fig.~1,
with the upper and lower lines corresponding to dominance by
the matrix elements $\langle {\cal O}^{\psi}_8(^1S_0) \rangle$
and $\langle {\cal O}^{\psi}_8(^3P_0) \rangle$, respectively.
This prediction stands in sharp contrast to the prediction of the
color-singlet model that $A(E)$ should approach $-0.85$
in the endpoint region.

Higher order radiative corrections and relativistic corrections will
produce a smooth transition between the moderate $z$ region where the
color-singlet contributions dominate and the endpoint region where the
color-octet contributions dominate.
Relativistic corrections to the color-singlet contribution
can  be taken into account by including additional
matrix elements in the general factorization formula (\ref{fact}).
At order $\alpha_s^2$, the first relativistic correction
will change the cross section by an amount of order $v^2$.  This could easily
change the color-singlet prediction for the asymmetry parameter $A$ at the
endpoint by about 30\%, but it is unlikely to change the sign.
At order $\alpha_s^3$, the relativistic correction diverges logarithmically
as $E \to E_{\rm max}$.  The singularity at the endpoint can be factored
into one of the color-octet matrix elements and it has therefore already been
taken into account in our analysis.

For a more quantitative analysis of $A(E)$, we require
phenomenological determinations of
$\langle {\cal O}^{\psi}_8(^1S_0) \rangle$ and
$\langle {\cal O}^{\psi}_8(^3P_0) \rangle$.  These matrix elements
are independent of the color-octet matrix element
$\langle {\cal O}^{\psi}_8(^3S_1) \rangle$ that plays an important
role in gluon fragmentation into $\psi$.  One process which provides
information on $\langle {\cal O}^{\psi}_8(^1S_0) \rangle$
and $\langle {\cal O}^{\psi}_8(^3P_0) \rangle$ is photoproduction.
The cross section for $\gamma N \to \psi + X$ was first calculated
in the color-singlet model by Berger and Jones \cite{berger}.
The relevant parton process at leading order in $\alpha_s$
is $\gamma g \to c \bar c g$.  As emphasized by Berger and Jones,
the color-singlet model breaks down in the endpoint region $z \to 1$,
where $z = E_\psi/E_\gamma$ is the ratio of the energies of the
$\psi$ and the photon in the nucleon rest frame.
{}From the point of view of the factorization approach of Ref.~\cite{BBL},
this breakdown of the color-singlet model can be attributed to
contributions from color-octet processes in the endpoint region $z \to 1$.
At leading order in $\alpha_s$, the relevant parton process is
$\gamma g \to c \bar c$, which produces color-octet $c \bar c$ pairs
in the endpoint region with angular momentum quantum numbers $^1S_0$
and $^3P_J$.  Thus it might be possible to use
data on photoproduction in the nearly elastic region $z \to 1$
to constrain the matrix elements $\langle {\cal O}^{\psi}_8(^1S_0) \rangle$
and $\langle {\cal O}^{\psi}_8(^3P_J) \rangle$.
Unfortunately, the endpoint $z=1$ for photoproduction is
dominated by elastic and diffractive $\psi$'s, which cannot be described
adequately within our factorization framework. The color-octet mechanism
produces inelastic $\psi$'s with $z$ near 1,
and it is not clear whether these can be separated from
the diffractive contribution.  In order to eliminate the diffractive
contribution, it may be necessary to go to higher momentum transfer and
determine the color-octet matrix elements from electroproduction data.

We have shown that the inclusive production cross section for
$J/\psi$ in $e^+e^-$ annihilation may be dominated by color-octet
production mechanisms near the upper endpoint of the $\psi$
energy spectrum.  The signal for the color-octet contributions
is a dramatic change in the angular distribution of the $\psi$,
which has the general form $1 + A(E) \cos^2 \theta$.
The prediction of the color-singlet model is $A \approx -0.84$
at the upper endpoint $E_{\rm max}$.
If the endpoint region is dominated by color-octet processes,
then $A(E)$ should change sign near the endpoint,
approaching a value $A \ge +0.62$.
Thus the angular distribution provides a simple and dramatic
signature for color-octet production mechanisms.
The enormous statistics available at CLEO makes it possible to
measure the angular distribution as a function of the momentum
$p_\psi$ of the $\psi$.
Preliminary results from CLEO indicate that the angular distribution
is consistent with the color-singlet model at moderate values of
$p_\psi$, but that the asymmetry parameter $A$ changes sign at the largest
values of $p_\psi$ \cite{klaus}
in agreement with the prediction from color-octet dominance.
Such a result would provide strong evidence that
color-octet processes play a significant role in charmonium  production.

This work was supported in part by the U.S.
Department of Energy, Division of High Energy Physics, under
Grant DE-FG02-91-ER40684.

\vfill\eject



\figure{
Angular distribution parameter $A$ as a function of the
momentum $p_\psi$ of the $\psi$ for $e^+ e^-$ annihilation
at $\sqrt{s} = 10.6 \; {\rm GeV}$.
The solid line is the prediction of the color-singlet model.
The hatched region is the band of predictions in the endpoint region,
assuming that this region is dominated by color-octet
production mechanisms.}

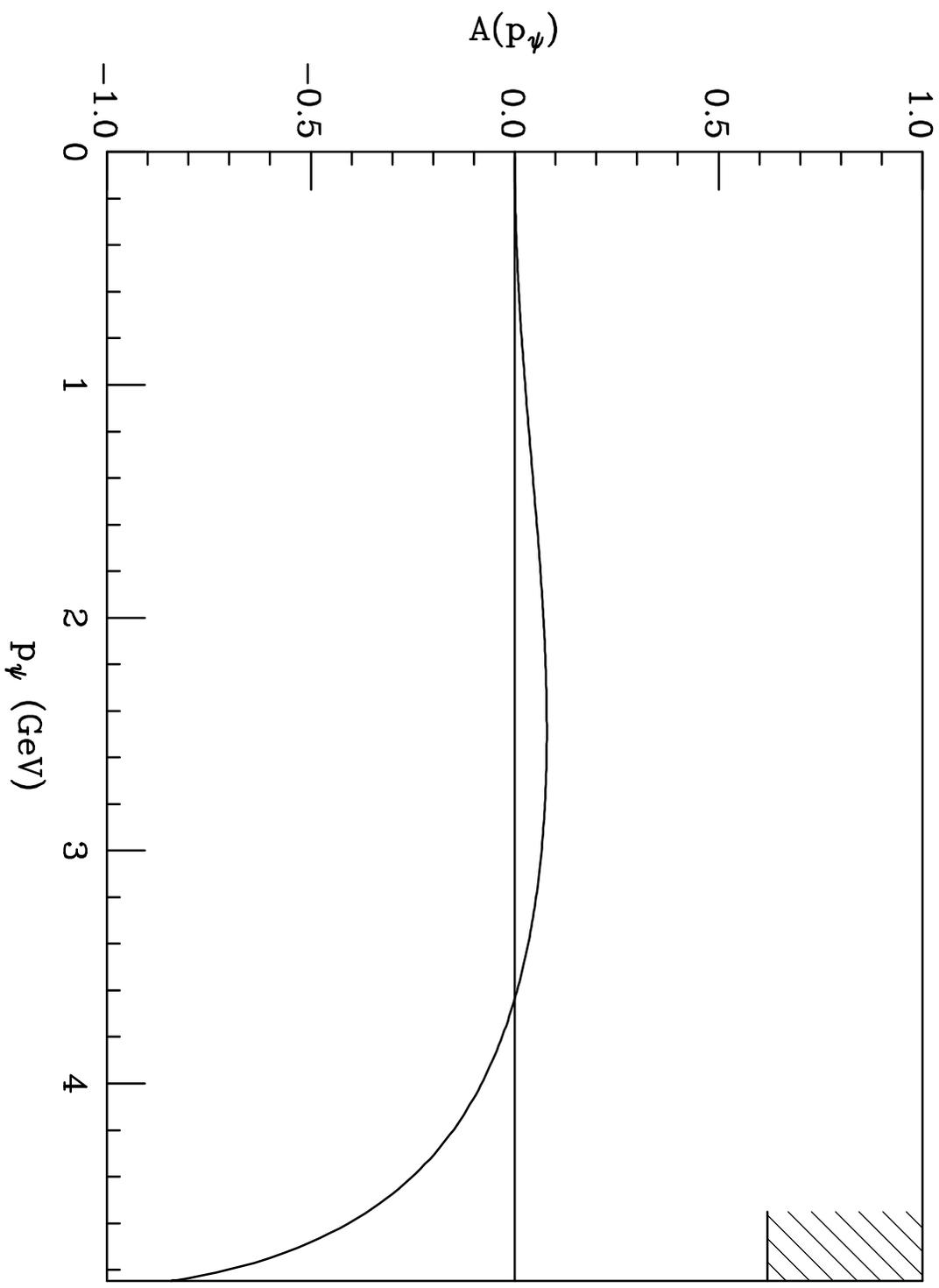
\end{document}